\pdfoutput=1
\documentclass{appolb}
\usepackage{cite}
\usepackage{amsmath}
\usepackage[colorlinks=true, urlcolor=blue, citecolor=blue, anchorcolor=blue]{hyperref}
\usepackage{stackengine}

\usepackage{graphicx}
\usepackage{caption}
\usepackage{float}

\begin{document}
\title{Nuclear Parton Distributions from Neural Networks%
\thanks{Talk presented by R.~A.~K. at Diffraction and Low-x 2018 conference.}%
}
\author{\textbf{The NNPDF Collaboration}:
\\Rabah Abdul Khalek, Jacob J. Ethier, and Juan Rojo
\address{Department of Physics and Astronomy,\\
Vrije Universiteit Amsterdam, 1081 HV
  Amsterdam, \\
  Nikhef Theory Group, Science Park 105, 1098 XG Amsterdam, The Netherlands.}
}
\maketitle
\begin{abstract}
In this contribution we present a status report on the recent progress towards an
analysis of nuclear parton distribution functions (nPDFs) using the NNPDF methodology. 
We discuss how the NNPDF fitting approach can be extended to account for the 
dependence on the atomic mass number $A$, and introduce novel algorithms to 
improve the training of the neural network parameters within the NNPDF framework. 
Finally, we present preliminary results of a nPDF fit to neutral current 
deep-inelastic lepton-nucleus scattering data, and demonstrate how one can validate 
the new fitting methodology by means of closure tests.
\end{abstract}
\PACS{13.60.-r\\}

\noindent
    \textbf{Introduction.}
Parton distribution functions (PDFs) are universal, process-independent objects 
describing the longitudinal motion of quarks and gluons within 
hadrons~\cite{Butterworth:2015oua,Gao:2017yyd}.
Since PDFs are difficult to compute from first principles, they 
are instead extracted from experimental data by means of a global analysis in the 
framework of QCD collinear factorization theorems.
Currently, the PDFs of nucleons bound within heavy nuclei 
(nPDFs)~\cite{Paukkunen:2018kmm} are less well understood than their free-nucleon 
counterparts, due primarily to the limited experimental constraints available.

This state of affairs is unfortunate,
since the determination of nPDFs is important to reveal the origin and properties of 
phenomena such as the Fermi motion, the EMC effect, nuclear shadowing, and possible
non-linear evolution effects in nuclei.
In addition, nPDFs are key inputs for the interpretation of heavy-ion collisions and the 
characterization of the Quark-Gluon Plasma (QGP), as well as for high-energy 
astrophysics such as theoretical predictions of neutrino-nucleus interaction 
cross-sections~\cite{Bertone:2018dse}.

Several groups have presented nPDF determinations in recent years.
Two of such analyses are
EPPS16~\cite{Eskola:2016oht}, which fit a nuclear modification factor with respect to
 the CT14~\cite{Dulat:2015mca} proton baseline, and nCTEQ15~\cite{Kovarik:2015cma}, 
which fit directly the nPDF shape by mimicking the parameterization used in the CTEQ 
proton fits~\cite{Pumplin:2002vw}.
This recent activity in global nPDF studies has been largely prompted by the availability of 
proton-lead collision observables such as dijet, $D$ meson, or $W$ 
and $Z$ gauge boson production. As in the case of proton PDFs, these measurements 
offer the potential of a greatly improved understanding of nPDFs and their uncertainties.\\

\noindent
    \textbf{Towards nNNPDF1.0.}
Following the NNPDF methodology~\cite{Ball:2008by,Ball:2012cx,Ball:2014uwa}
(see~\cite{Rojo:2018qdd} for a summary), we adopt here artificial neural networks
(ANNs) as universal unbiased interpolants to parameterize the $x$ and $A$ dependence 
of the nPDFs. 
As an initial study, we consider only observables from neutral current (NC) 
deep-inelastic scattering (DIS) off heavy nuclei, which are assumed to be isoscalar.

The description of isoscalar nuclei observables in DIS below the $Z$-boson pole 
requires the parametrisation of three independent nPDFs, which are taken to be the 
quark singlet $\Sigma$, the gluon $g$, and the quark non-singlet octet $T_8$ distributions.
In this basis, for example, the NC DIS structure function $F_2^A$ is given by
\begin{align}
\label{eq:f2}
    F_2^A(x,Q^2) & = \Gamma_{2,\Sigma}^S(x,Q_0^2,Q^2)\otimes \Sigma(x,A,Q_0^2) \nonumber \\
    & + \Gamma_{2,g}^{\rm S}(x,Q_0^2,Q^2) \otimes g(x,A,Q_0^2) \\
    & + \Gamma_{2,T_8}^{\rm NS}(x,Q_0^2,Q^2) \otimes T_8(x,A,Q_0^2) \nonumber \, ,
\end{align}
where the $\Gamma$ factors encode both the hard--scattering coefficient functions
and the DGLAP evolution kernels.
The nPDFs are then parametrised at an initial scale denoted by $Q_0$, and depend 
both on the partonic momentum fraction $x$ and the mass number $A$.

Following Ref.~\cite{Bertone:2016lga}, the convolutions in Eq.~(\ref{eq:f2}) can be 
reduced to a scalar product by means of an expansion over a set of interpolating 
polynomials, allowing us to write
\begin{align}
\label{eq:f2_2}
F_2^A(x,Q^2) &= \sum_i^{n_f} \sum_{\alpha}^{n_x} \widetilde{\Gamma}_{i,\alpha}(x,x_{\alpha},Q^2,Q^2_0)
\cdot q_i(x_{\alpha},A,Q_0^2) 
\end{align}
where $\widetilde{\Gamma}$ stand for the precomputed {\tt FastKernel} grids that 
contain all the perturbative information relevant for the calculation of $F_2^A$, and 
$q_i(x,A,Q_0^2)$ represent the initial scale nPDF for the flavour $i$ in a given basis.
Note that in Eq.~(\ref{eq:f2_2}) only the values of the input PDFs at a finite $n_x$-sized
grid are required to compute the structure functions, leading to a significant improvement 
of the numerical computation with respect to the convolutions in Eq.~(\ref{eq:f2}).

The three independent nPDFs that enter Eq.~(\ref{eq:f2_2}) are parametrised as
\begin{itemize}
\item $\Sigma(x,A,Q_0)=(1-x)^{\alpha_{\Sigma}}x^{-\beta_\Sigma}{\rm NN}_{\Sigma}(x,A)$  ,
\item $g(x,A,Q_0)=A_g(1-x)^{\alpha_{G}}x^{-\beta_g}{\rm NN}_{g}(x,A)$ ,
\item $T_8(x,A,Q_0)=(1-x)^{\alpha_{T_8}}x^{-\beta_{T_8}}{\rm NN}_{T_8}(x,A)$ ,
\end{itemize}
where ${\rm NN}_i$ corresponds to the output of the ANN for a given flavor $i$. 
The preprocessing exponents~\cite{Ball:2016spl} $\alpha_i$ and $\beta_i$ facilitate the 
training procedure and can either be fitted or drawn at random from a range determined 
iteratively.
Furthermore, we fix the overall normalisation of the gluon nPDF,
\begin{equation}
A_g \equiv \Bigg(1-\int_0^1\Sigma(x,A,Q_0)dx\Bigg)\Bigg/\int_0^1g(x,A,Q_0)dx
\end{equation}
so that the momentum sum rule is satisfied. In general, this normalisation is different 
for every value of $A$.

\begin{figure}[t]
\centerline{%
\includegraphics[width=0.8\textwidth]{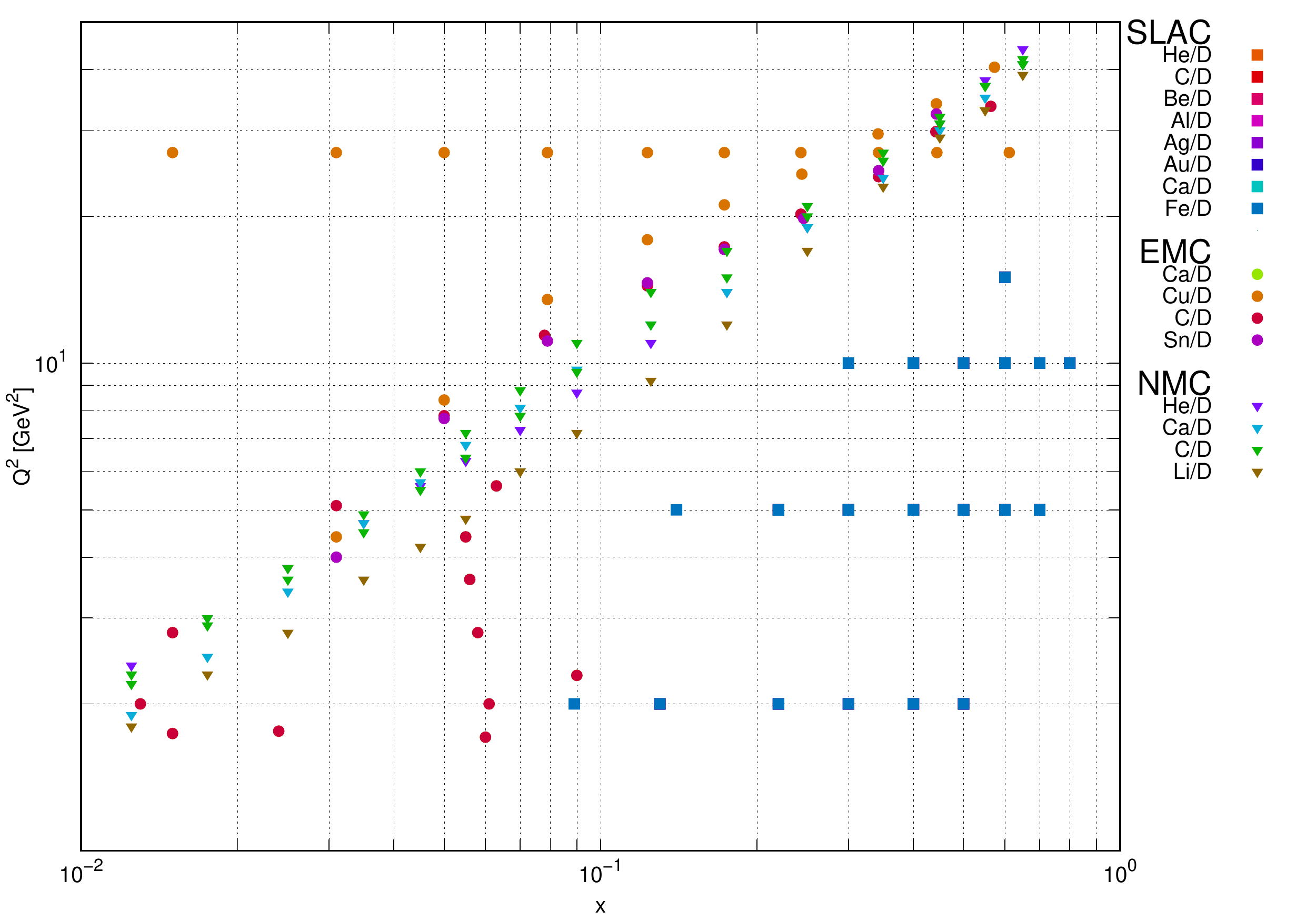}}
\caption{\label{fig:kinplot} The kinematic coverage in $x$ and $Q^2$ 
of the NC DIS data included in this work.}
\end{figure}

Concerning the input dataset, we consider here a similar set of nuclear NC DIS 
measurements that were used by EPPS16 and nCTEQ15.
In Fig.~\ref{fig:kinplot} the kinematic coverage of the $(x,Q^2)$ plane of the 
nuclear DIS data are shown.
Here, the coverage in $x$ is significantly reduced compared to the 
proton case ($x > 10^{-2}$ versus $x > 10^{-5}$ respectively).
Enlarging this kinematic range to smaller values of $x$ and higher values of
$Q^2$ is possible by means of the RHIC and LHC data on nucleon-nucleus 
collisions.\\

\noindent
    \textbf{Neural network training.}
The underlying procedure for any optimisation problem, such as the 
present one, can be summarized by
    \begin{equation}
      \label{eq:optimisation}
    \stackunder{min}{$\pmb \omega$}\,\,C(f({\pmb \omega})),
\end{equation}
where $C$ is a cost function to minimize and $f$ is the target function that depends
on a vector of parameters ${\pmb \omega}$.
In our case, the target functions that need to be determined from
the data are the nPDFs parametrised by the neural networks, while ${\pmb \omega}$ 
represents the neural network weights and thresholds. The cost function is defined 
here to be the $\chi^2$, which measures the agreement between the experimental 
data points $D_i$ and the corresponding theoretical predictions $T_i$ of nuclear 
DIS observables, 

\begin{equation}
  \chi^{2}=\sum_{i=1}^{n_{\rm dat}} \frac{(T_i[{f({\pmb \omega})}] - D_i)^2}{
    \sigma^{2}_i} \, ,
\end{equation}
where $\sigma_i$ is the total experimental statistical and systematic uncertainties
added in quadrature.

There are different options that can be used to solve Eq.~(\ref{eq:optimisation}).
Previous NNPDF global fits have been based either
in Genetic Algorithms (GAs) or
the Covariance Matrix Adaptation - Evolutionary Strategy (CMA-ES) algorithms.
Both methods require knowledge only on the local values of the $\chi^2$ and not
of its derivatives.
Here for the first time in the context of NNPDF studies we have
implemented the method of gradient descent, one of the most widely used 
minimization techniques in machine learning applications.
In this procedure, the parameters are shifted by an amount
proportional to the negative of the gradient of the cost function evaluated at the 
current position in parameter space,
\begin{equation}
  \label{sec:iterate}
  \omega_{i} \to \omega_{i} - \frac{ \eta }{n_{\rm par}} \frac{\partial
    \chi^{2}}{\partial \omega_{i}} \, ,
\end{equation}
where $\omega_{i}$ is one of the $n_{\rm par}$ free parameters of the ANN
and $\eta$ is a hyperparameter of the algorithm known as the learning rate.
In this work, the gradients are computed numerically by means of 
automatic differentiation using the {\tt TensorFlow} library~\cite{tensorflow2015-whitepaper}.
 The updating process in Eq.~(\ref{sec:iterate}) is then iterated until a suitable set of
 convergence criteria is satisfied, for instance, using look-back or early stopping
 with cross-validation.\\

\begin{figure}[t]
\centerline{
\includegraphics[width=1.2\textwidth]{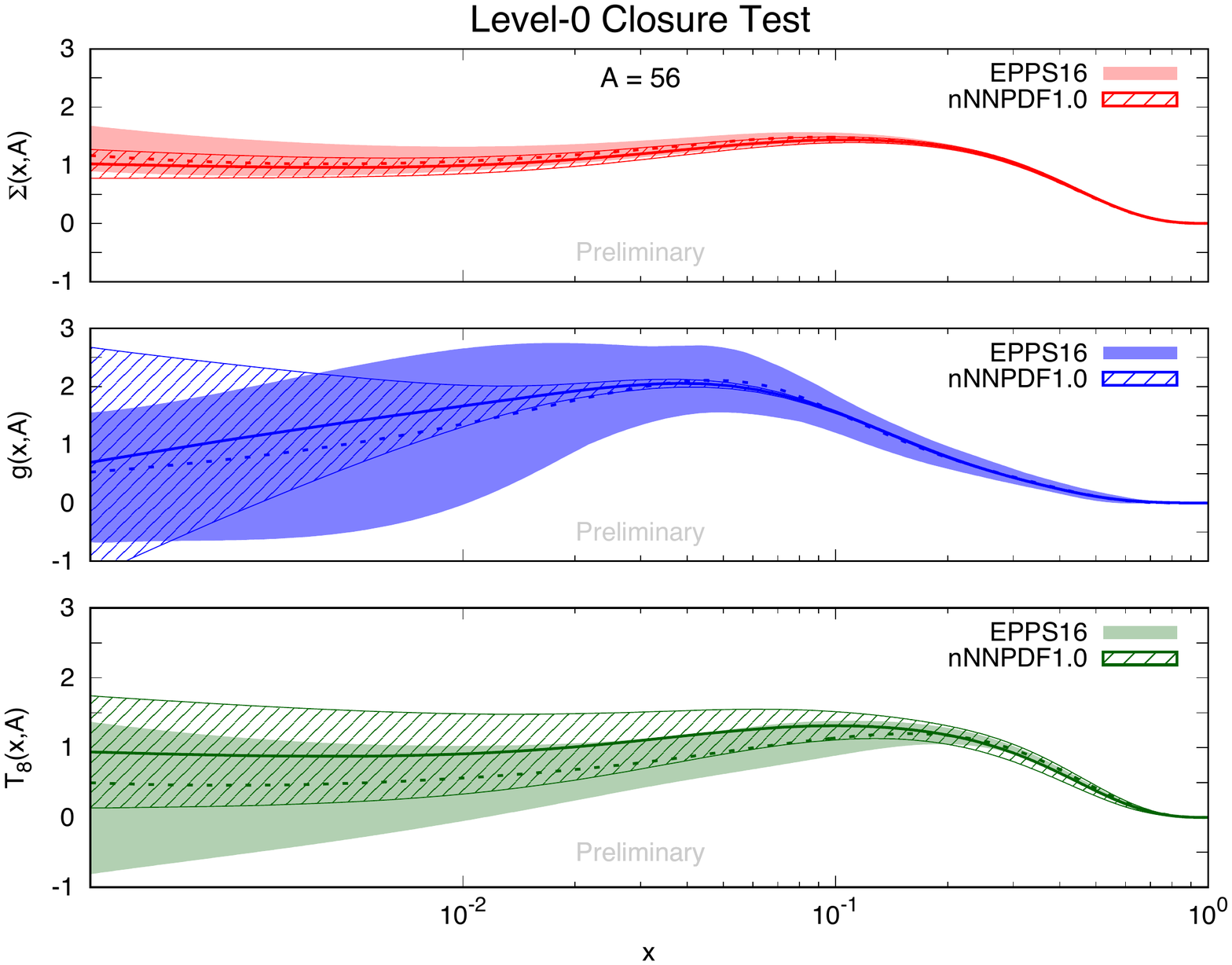}}
\caption{\small The results of the nNNPDF1.0 Level 0 closure test. The nNNPDF1.0
result (solid line with hashed bands) is compared to the EPPS16 nPDF result
(solid line with shaded bands) for the iron nucleus ($A=56$) at $Q_0 = 1.3$ GeV 
for the quark singlet $\Sigma$ (red), the gluon $g$ (blue), and the quark non-singlet 
octet distribution $T_8$ (green). 
 \label{fig:A56} }
\end{figure}

\noindent
 \textbf{Preliminary results.}
 The application of the NNPDF methodology to a QCD analysis of nuclear
 parton distributions can be validated by means of closure tests, as was done
 in previous global fits of proton PDFs~\cite{Ball:2014uwa} and fragmentation
 functions~\cite{Bertone:2017tyb}.
 In these closure tests, pseudo-data is generated based on an established
 theoretical input. In this work, we construct pseudo-data with Eq.~(\ref{eq:f2}) 
 up to NLO in perturbative QCD using the EPPS16 nPDF set.
 A fit is then performed to this pseudo-data, and by comparing the fit
 output to the known input we can assess if a given fitting methodology
 is working successfully.
 Since this pseudo-data is free from possible data inconsistencies or limitations
 in the theory calculations, it provides a clean testing group to validate
 the fitting strategy employed.

\begin{figure}[t]
\centerline{%
\includegraphics[width=1.2\textwidth,keepaspectratio]{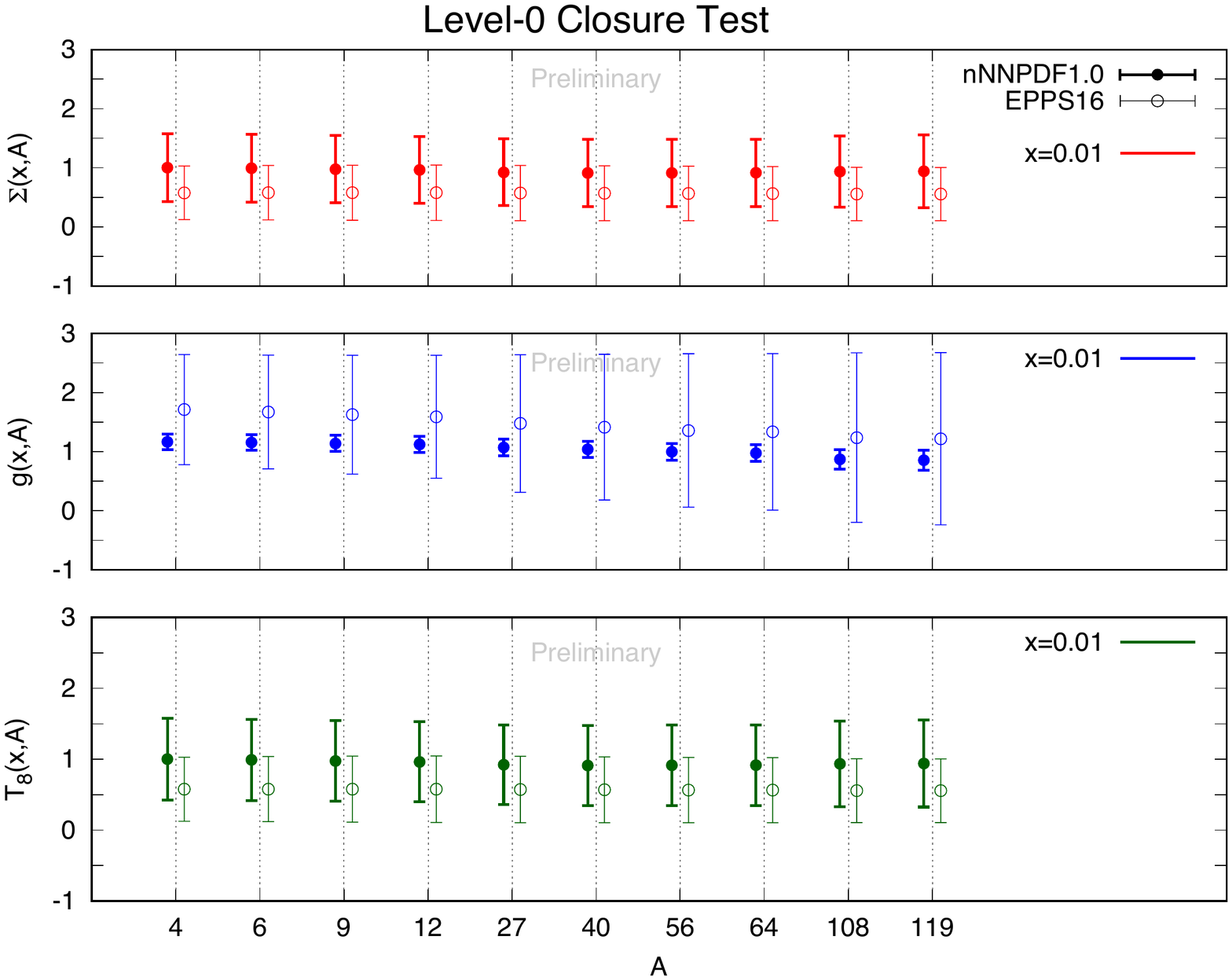}}
\caption{\small Similar to Fig.~\ref{fig:A56}, but now showing the comparison
for discrete values of mass number $A$ at fixed $x=0.01$.  
\label{fig:Adep}}
\end{figure}

 We have performed the simplest type of closure test, known as Level 0 (L0),
 where the pseudo-data coincides with the EPPS16 prediction without any additional
 set of statistical noise added.
 In this case, a successful fit should be able to reach asymptotically $\chi^2\to 0$,
 so that the fit predictions at the structure function level are identical
 to those obtained with the input EPPS16 theory.
 For this L0 closure test, we have used a single neural network
 with three input nodes $(x,\ln x,A)$, three output nodes ${\rm NN}_{\Sigma}$,
 ${\rm NN}_{g}$, ${\rm NN}_{T_8}$, and a single hidden layer with 20 neurons.
 This ANN architecture contains then 143 free parameters.
 Our final result consists of 200 Monte Carlo ``replicas'', which in L0 tests correspond 
 to different initial parameter values that are randomly chosen for each fit.

 In Fig.~\ref{fig:A56} we display the $x$-dependence of the
 preliminary nNNPDF1.0 L0 closure test fit results for iron nuclei ($A=56$).
 Here the gluon, quark singlet $\Sigma$, and non-singlet octet $T_8$ distributions
 are shown at the input scale $Q_0=1.3$ GeV and are compared with the EPPS16 result.
 While the EPPS16 uncertainty band is computed using the asymmetric Hessian method, the 
 nNNPDF1.0 band instead represents the variance evaluated over the 200 fitted replicas.
 The fact that both central values agree reasonably well, especially for the gluon,
 is a first indication that the closure test is successful.
 Note that $\Sigma$ and $T_8$ are strongly anti-correlated in the data region, since
 the actual quantity which is being constrained from data is $F_2^A \propto \Sigma+T_8/4$.
 Also, for L0 closure tests the two error bands have different statistical
 interpretations and cannot be compared directly.

 In Fig.~\ref{fig:Adep} we present a similar comparison as in
 Fig.~\ref{fig:A56}, but now show the mass number
 $A$ dependence for a fixed value of $x$.
 Here we see that the closure test results agree with the input EPPS16 theory within
 uncertainties, and that they reproduce the same qualitative behavior as $A$ is varied.\\

\noindent
\textbf{Next steps.}
In this contribution we have presented the initial steps towards
the first determination of the nuclear parton distributions in the framework
of the NNPDF methodology.
We have validated the effectiveness of improved neural network training algorithms,
in particular the gradient descent minimization with {\tt TensorFlow}.
At the closure test level, we have shown that we are able to reproduce the results
of the chosen input theory, in this case EPPS16 nPDFs.
Work in progress is now focused on extending our approach to Level 1 and 2 closure tests,
as well as to a full QCD analysis of experimental data.
Following an initial study on DIS measurements, we aim to deliver a full-fledged 
global nPDF fit that accounts for all available experimental constraints and
is based on state-of-the-art theoretical calculations.\\

\noindent
\textbf{Acknowledgements.}
This research has been supported 
by a European Research Council Starting grant ``PDF4BSM", and
by the Netherlands Organization for Scientific
Research (NWO).

\bibliography{low-x}

\providecommand{\href}[2]{#2}\begingroup\raggedright\begin{thebibliography}{10}

\bibitem{Butterworth:2015oua}
J.~Butterworth et~al., {\it {PDF4LHC recommendations for LHC Run II}},  {\em J.
  Phys.} {\bf G43} (2016) 023001, [\href{http://arxiv.org/abs/1510.03865}{{\tt
  arXiv:1510.03865}}].

\bibitem{Gao:2017yyd}
J.~Gao, L.~Harland-Lang, and J.~Rojo, {\it {The Structure of the Proton in the
  LHC Precision Era}},  {\em Phys. Rept.} {\bf 742} (2018) 1--121,
  [\href{http://arxiv.org/abs/1709.04922}{{\tt arXiv:1709.04922}}].

\bibitem{Paukkunen:2018kmm}
H.~Paukkunen, {\it {Nuclear PDFs Today}},  in {\em {9th International
  Conference on Hard and Electromagnetic Probes of High-Energy Nuclear
  Collisions: Hard Probes 2018 (HP2018) Aix-Les-Bains, Savoie, France, October
  1-5, 2018}}, 2018.
\newblock \href{http://arxiv.org/abs/1811.01976}{{\tt arXiv:1811.01976}}.

\bibitem{Bertone:2018dse}
V.~Bertone, R.~Gauld, and J.~Rojo, {\it {Neutrino Telescopes as QCD
  Microscopes}},  \href{http://arxiv.org/abs/1808.02034}{{\tt
  arXiv:1808.02034}}.

\bibitem{Eskola:2016oht}
K.~J. Eskola, P.~Paakkinen, H.~Paukkunen, and C.~A. Salgado, {\it {EPPS16:
  Nuclear parton distributions with LHC data}},  {\em Eur. Phys. J.} {\bf C77}
  (2017), no.~3 163, [\href{http://arxiv.org/abs/1612.05741}{{\tt
  arXiv:1612.05741}}].

\bibitem{Dulat:2015mca}
S.~Dulat, T.-J. Hou, J.~Gao, M.~Guzzi, J.~Huston, P.~Nadolsky, J.~Pumplin,
  C.~Schmidt, D.~Stump, and C.~P. Yuan, {\it {New parton distribution functions
  from a global analysis of quantum chromodynamics}},  {\em Phys. Rev.} {\bf
  D93} (2016), no.~3 033006, [\href{http://arxiv.org/abs/1506.07443}{{\tt
  arXiv:1506.07443}}].

\bibitem{Kovarik:2015cma}
K.~Kovarik et~al., {\it {nCTEQ15 - Global analysis of nuclear parton
  distributions with uncertainties in the CTEQ framework}},  {\em Phys. Rev.}
  {\bf D93} (2016), no.~8 085037, [\href{http://arxiv.org/abs/1509.00792}{{\tt
  arXiv:1509.00792}}].

\bibitem{Pumplin:2002vw}
J.~Pumplin, D.~R. Stump, J.~Huston, H.~L. Lai, P.~M. Nadolsky, and W.~K. Tung,
  {\it {New generation of parton distributions with uncertainties from global
  QCD analysis}},  {\em JHEP} {\bf 07} (2002) 012,
  [\href{http://arxiv.org/abs/hep-ph/0201195}{{\tt hep-ph/0201195}}].

\bibitem{Ball:2008by}
{\bf NNPDF} Collaboration, R.~D. Ball, L.~Del~Debbio, S.~Forte, A.~Guffanti,
  J.~I. Latorre, A.~Piccione, J.~Rojo, and M.~Ubiali, {\it {A Determination of
  parton distributions with faithful uncertainty estimation}},  {\em Nucl.
  Phys.} {\bf B809} (2009) 1--63, [\href{http://arxiv.org/abs/0808.1231}{{\tt
  arXiv:0808.1231}}]. [Erratum: Nucl. Phys.B816,293(2009)].

\bibitem{Ball:2012cx}
R.~D. Ball et~al., {\it {Parton distributions with LHC data}},  {\em Nucl.
  Phys.} {\bf B867} (2013) 244--289,
  [\href{http://arxiv.org/abs/1207.1303}{{\tt arXiv:1207.1303}}].

\bibitem{Ball:2014uwa}
{\bf NNPDF} Collaboration, R.~D. Ball et~al., {\it {Parton distributions for
  the LHC Run II}},  {\em JHEP} {\bf 04} (2015) 040,
  [\href{http://arxiv.org/abs/1410.8849}{{\tt arXiv:1410.8849}}].

\bibitem{Rojo:2018qdd}
J.~Rojo, {\it {Machine Learning tools for global PDF fits}},  in {\em {13th
  Conference on Quark Confinement and the Hadron Spectrum (Confinement XIII)
  Maynooth, Ireland, July 31-August 6, 2018}}, 2018.
\newblock \href{http://arxiv.org/abs/1809.04392}{{\tt arXiv:1809.04392}}.

\bibitem{Bertone:2016lga}
V.~Bertone, S.~Carrazza, and N.~P. Hartland, {\it {APFELgrid: a high
  performance tool for parton density determinations}},  {\em Comput. Phys.
  Commun.} {\bf 212} (2017) 205--209,
  [\href{http://arxiv.org/abs/1605.02070}{{\tt arXiv:1605.02070}}].

\bibitem{Ball:2016spl}
R.~D. Ball, E.~R. Nocera, and J.~Rojo, {\it {The asymptotic behaviour of parton
  distributions at small and large $x$}},  {\em Eur. Phys. J.} {\bf C76}
  (2016), no.~7 383, [\href{http://arxiv.org/abs/1604.00024}{{\tt
  arXiv:1604.00024}}].

\bibitem{tensorflow2015-whitepaper}
M.~Abadi, A.~Agarwal, P.~Barham, E.~Brevdo, Z.~Chen, C.~Citro, G.~S. Corrado,
  A.~Davis, J.~Dean, M.~Devin, S.~Ghemawat, I.~Goodfellow, A.~Harp, G.~Irving,
  M.~Isard, Y.~Jia, R.~Jozefowicz, L.~Kaiser, M.~Kudlur, J.~Levenberg,
  D.~Man\'{e}, R.~Monga, S.~Moore, D.~Murray, C.~Olah, M.~Schuster, J.~Shlens,
  B.~Steiner, I.~Sutskever, K.~Talwar, P.~Tucker, V.~Vanhoucke, V.~Vasudevan,
  F.~Vi\'{e}gas, O.~Vinyals, P.~Warden, M.~Wattenberg, M.~Wicke, Y.~Yu, and
  X.~Zheng, {\it {TensorFlow}: Large-scale machine learning on heterogeneous
  systems},  2015.
\newblock Software available from tensorflow.org.

\bibitem{Bertone:2017tyb}
{\bf NNPDF} Collaboration, V.~Bertone, S.~Carrazza, N.~P. Hartland, E.~R.
  Nocera, and J.~Rojo, {\it {A determination of the fragmentation functions of
  pions, kaons, and protons with faithful uncertainties}},  {\em Eur. Phys. J.}
  {\bf C77} (2017), no.~8 516, [\href{http://arxiv.org/abs/1706.07049}{{\tt
  arXiv:1706.07049}}].

\end{thebibliography}\endgroup

\end{document}